\documentclass[aip,graphicx, amsmath,amssymb,reprint]{revtex4-1}

\usepackage[version=3]{mhchem}
\usepackage{graphicx}
\usepackage{dcolumn}
\usepackage{bm}
\usepackage[warning,debug,nosepfour,autolanguage]{numprint}

\usepackage[utf8]{inputenc}
\usepackage[T1]{fontenc}
\usepackage{mathptmx}
\usepackage{comment}

\draft 

\newcommand{\figref}[1]{Fig.~\ref{#1}}
\newcommand{\bfigref}[1]{Figure~\ref{#1}}

\begin{document}


\title{Etching process of narrow wire and application to tunable-barrier electron pump} 


\author{Shota Norimoto}
\email[]{shota.norimoto2@oist.jp}
\affiliation{Graduate School of Science, Osaka University, Toyonaka, Osaka 560-0043, Japan}

\author{Shuichi Iwakiri}
\affiliation{Graduate School of Science, Osaka University, Toyonaka, Osaka 560-0043, Japan}

\author{Masahiko Yokoi}
\affiliation{Graduate School of Science, Osaka University, Toyonaka, Osaka 560-0043, Japan}

\author{Tomonori Arakawa}
\affiliation{Graduate School of Science, Osaka University, Toyonaka, Osaka 560-0043, Japan}
\affiliation{Center for Spintronics Research Network, Graduate School of Engineering Science, Osaka University, Toyonaka, Osaka 560-8531, Japan}

\author{Yasuhiro Niimi}
\affiliation{Graduate School of Science, Osaka University, Toyonaka, Osaka 560-0043, Japan}
\affiliation{Center for Spintronics Research Network, Graduate School of Engineering Science, Osaka University, Toyonaka, Osaka 560-8531, Japan}

\author{Kensuke Kobayashi}
\affiliation{Graduate School of Science, Osaka University, Toyonaka, Osaka 560-0043, Japan}
\affiliation{Institute for Physics of Intelligence and Department of Physics, The University of Tokyo, Bunkyo-ku, Tokyo 113-0033, Japan}


\date{\today}

\begin{abstract}
Single electron sources have been studied as a device to establish an electric current standard for 30 years and recently as an on-demand coherent source for Fermion quantum optics.
In order to construct the single electron source on a \ce{GaAs/AlGaAs} two-dimensional electron gas (2DEG), it is often necessary to fabricate a sub-micron wire by etching.
We have established techniques to make the wire made of the fragile 2DEG by combining photolithography and electron beam lithography with one-step photoresist coating, which enables us to etch fine and coarse structures simultaneously.
It has been demonstrated that a single electron source fabricated on the narrow wire pumps fixed number of electrons per one cycle with radio frequency.
The fabrication technique improves the lithography process with lower risk to damage the 2DEG and is applicable to etching of other materials and dry etching.
\end{abstract}


\maketitle 


Single electron sources have a long research history~\cite{Kouwenhoven1991,Nagamune1994,Keller1996,Talyanskii1997,Fujiwara2004,Blumenthal2007,Feve2007,Kaestner2008PRB,Kaestner2008APL,pekola2013,Yamahata2014}.
Some of them are fabricated on a \ce{GaAs}-based two dimensional electron gas (2DEG) system which provides high controllability of electrons.
One of the big motivations to study single electron sources is to establish a quantum current standard with elementary charge $e$ and frequency $f$\cite{Kouwenhoven1991,pekola2013,Yamahata2014}.
At the ``26e Conf$\acute{\textrm{e}}$rence G$\acute{\textrm{e}}$n$\acute{\textrm{e}}$rale des Poids et Mesures'' held in the fall of 2018, the revised definition of the SI basic units, including the definition of the electric current, was adopted, and the revised SI became effective on May 20, 2019 (World Metrology Day)~\cite{BureauInternationaldesPoidsetMeasures2018}.
The present practical electric current standard is, however, derived by the voltage standard defined by the Josephson effect and the resistance standard defined by the quantum Hall effect through Ohm's law, while ampere ($\textrm{A}$) is one of the seven SI base units~\cite{SI2019}.

In addition to the above motivation, the recent researches on single electron sources aim at developing on-demand coherent sources controlled by radio frequency (RF).
Because the timing and energy of the pumped electrons can be controlled, two-electron interference experiments and pump-probe measurements were demonstrated~\cite{Crowell2013,Freulon2015,Fletcher2013,Kataoka2016}.
The pump-probe measurement for the quantum Hall edge channel revealed the velocity of the pumped electron and their distributions in the time and energy space~\cite{Fletcher2013,Kataoka2016}.
Thus, the single electron source is a powerful tool to study the dynamics of electrons in quantum systems.

Most of the single electron sources utilize discrete levels in a quantum dot formed in sub-micron structures to operate single electrons.
Especially, a single electron source with quantum dot in a narrow 2DEG wire ``tunable-barrier electron pump'' in the quantum Hall regime has been an important experimental platform, as the edge state is dissipationless one-dimensional waveguide for \numprint[\mu m]{several tens}~\cite{Fletcher2013,Ubbelohde2015,Kataoka2016,Roulleau2008,Johnson2018}.
Regarding fabrication of a narrow 2DEG wire, on the other hand, there are several technical challenges in the fabrication process.
First, electron beam (EB) resists have weaker resistance against an etchant used to fabricate narrow 2DEG wires compared with photoresists.
Second, the etchant tends to intrude into the interface between the substrate and resist.

Here we present a technique to make a 2DEG narrow wire by combining the photolithography and EB lithography with one-step photoresist coating.
The main point is that the patterning of the wet-etching process is performed by only using a positive photoresist S1813G (Shipley), which has a strong resistance against acid.
Some positive photo resists work as a negative EB resist~\cite{Nakano2016}.
We perform EB lithography followed by photolithography after the coating of this photo resist.
When we perform the wet-etching process to fabricate a narrow sub-micron sized wire for single electron source devices, a special care is required so that the resist should adhere strongly to the wafer surface to possess sufficient resistance against a chemical etchant.
The device properties of the fabricated tunable-barrier electron pump is also be presented here.

\begin{figure}
    \centering
	\includegraphics[width = 7cm]{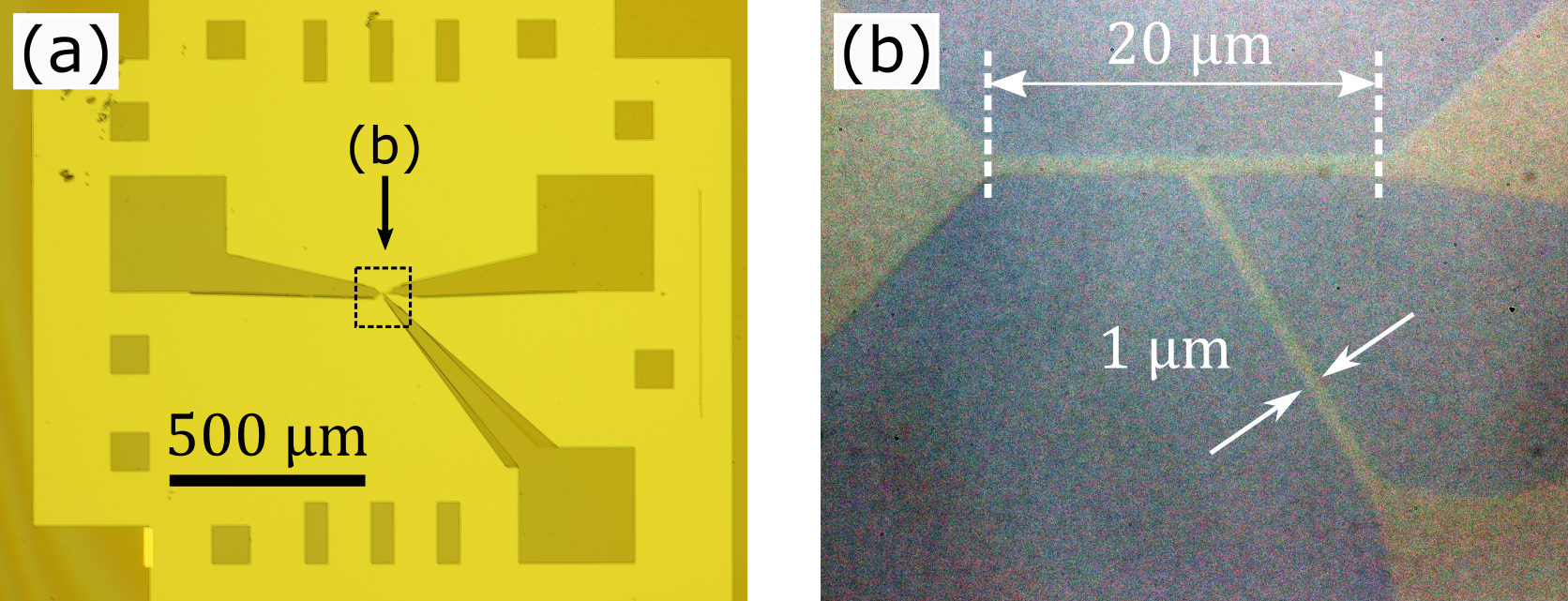}
	\caption{Optical photographs of the sample. (a) The image of the sample after development. The dark brown areas are the regions covered by the photoresist. (b) The image of the sample after the wet etching and the removal of the photoresist. In order to clarify the area of 2DEG, the picture is adjusted the contrast and brightness.
	}
	\label{nanowire}	
\end{figure}
The device are made from \ce{GaAs/AlGaAs} heterostructures with a 2DEG \numprint[nm]{75} below the surface.
The wafer was provided by Sumitomo Electric Industries, Ltd., whose electron density and mobility are \numprint[cm^{-2}]{2.52e11} and \numprint[cm^{2}/Vs]{1.77e5} at \numprint[K]{77}, respectively.

Before discussing the lithography process, the cleaning of the wafer is explained.
Generally, the cleaning process with acid or alkali is performed to improve adhesion of a resist to a substrate with peering an oxidized layer.
In this process, we also have to take care not to damage the 2DEG because our 2DEG substrate has only a \numprint[nm]{5} thickness capping \ce{i-GaAs} layer.
For this purpose, hexamethyldisilazane (HMDS) treatment is performed after cleaning the substrate with organic solvents (Remover PG, Acetone and 2-propanol) and ultraviolet ozone cleaner.
We expose the wafer to HMDS vapor in a draft chamber.
The wafer and a very small cup (less than \numprint[ml]{0.1} for \numprint[mm^{2}]{50}) filled with HMDS (AZ promoter, AZ products) are covered by a small beaker and are heated up to \numprint[\raisebox{1.0ex}{{\tiny$\circ$}}C]{135}.
After evaporation of the HMDS, the wafer should be kept for \numprint[min]{5} without the lid to dry it out.

The wafer is spin-coated with S1813G with slope \numprint[s]{5}/ \numprint[rpm]{5000} \numprint[s]{50}/ slope \numprint[s]{5} after cooling down the wafer.
The wafer is baked at \numprint[\raisebox{1.0ex}{{\tiny$\circ$}}C]{90} for \numprint[min]{10}.
Fine patterns are lithographed by \numprint[keV]{100} EB with the dose condition of \numprint[\mu C/cm ^{2}]{200-600} to draw \numprint[\mu m]{50-0.5} width-line by using Electron Beam Lithography System ELS7000 (ELIONIX INC.).
The area around the EB lithographed area has to be exposed by ultraviolet because it cannot be developed.
S1813G is developed by MF319 (Shipley) as performed in the conventional photolithography.
A large pattern lithographed by light and small pattern drawn by EB are developed at the same time, which is the most important point of the present fabrication technique.
This is an advantage of our method because it enables us to complete etching with one-step photoresist coating.
The development condition is to sink the substrate into MF319 for \numprint[s]{30} and to rinsed with water.
In order to improve the homogeneity of the development, the developer should be stirred.
Stirring chemicals prevents the reaction products from staying around the substrate so that they do not inhibit the expected chemical reaction.

\bfigref{nanowire}~(a) shows the developed pattern which contains areas lithographed by EB and ultraviolet.
In order to remove the residual resist, the wafer is cleaned with ultraviolet ozone cleaner (UV-1, SAMCO Inc.).
The 2DEG substrate is etched by an etchant which consists of \ce{H2SO4} (\numprint[wt\%]{96}): \ce{H2O2} (\numprint[wt\%]{30-35}): \ce{H2O} (distilled water) with a ratio of 3: 1: 200 in volume for \numprint[min]{1} with stirring the etchant.
The substrate should be rinsed by distilled water stream generated by a syringe in order to prevent contamination of the rinse and to stop etching completely.
\bfigref{nanowire}~(b) is an optical image of the etched fine structure after removal of the resist with hot (\numprint[\raisebox{1.0ex}{{\tiny$\circ$}}C]{60-90}) Remover PG.
In this way, a fine structure narrower than \numprint[\mu m]{1} is successfully fabricated.
The height of the mesa is around \numprint[nm]{30}.

\begin{figure}
	\centering
	\includegraphics[width=7cm]{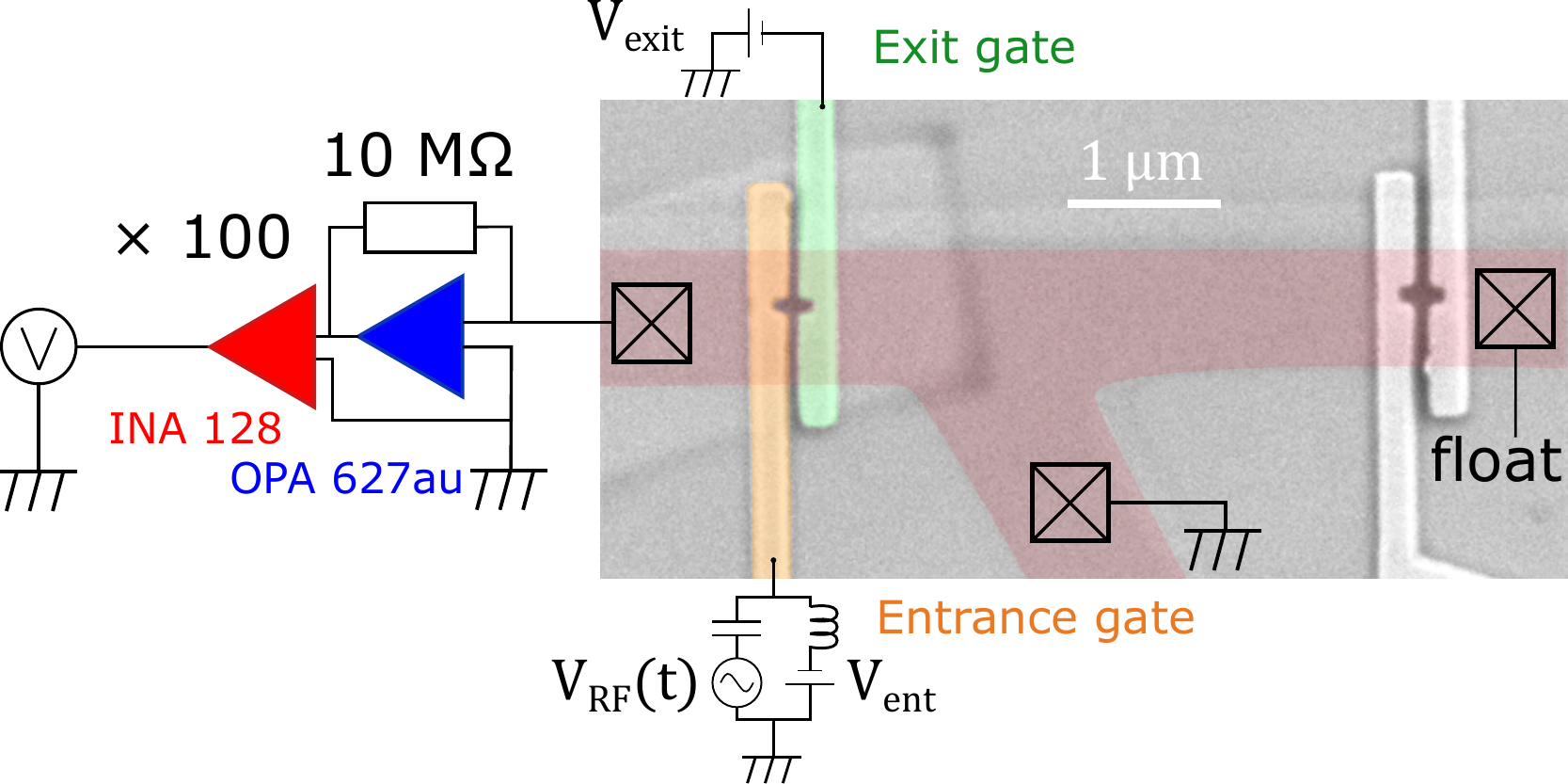}
	\caption{
False-colour SEM image of the sample and schematics of the circuits.
The 2DEG, Ohmic-contacts, entrance gate, and exit gate are indicated by red region, cross marks in boxes, orange region, and green region, respectively.
White regions in the right-hand side indicate gate electrodes, which also work as a pumping device.
}
	\label{SEM}	
\end{figure}

As the final process, the Ohmic contacts and the metallic gate electrodes are fabricated by the conventional fabrication techniques.
The right panel of \figref{SEM} shows a scanning electron microscope (SEM) image of the fabricated device.
The 2DEG is digitally assigned the false color of red.
Thin films of gold, germanium and nickel deposited on the substrate were alloyed at \numprint[\raisebox{1.0ex}{{\tiny$\circ$}}C]{440} to get the Ohmic contact between the surface and the 2DEG.
The metallic gate electrodes (orange, green and white regions in \figref{SEM}) consists of \ce{Ti} \numprint[nm]{5}/ \ce{Au} \numprint[nm]{100}.

\bfigref{SEM} shows the measurement setup for single electron pumping.
Each pair of two gate electrodes on the narrow 2DEG wire (the orange and green ones and the whites ones) defines a tunable-barrier electron pump, which enables us to operate a single or several electrons by applying RF signals on one gate electrode\cite{Fujiwara2008,Kaestner2008PRB,Kaestner2015}.
The number of pumped electrons in one cycle of RF signal is tuned by applying additional DC voltages to the two gate electrodes.
Schematics of the pumping process of tunable-barrier electron source were shown in Refs.[22, 23].
The present device, namely a single electron source that has two drains and gate electrodes to reflect hot electrons, is designed for time-resolved spectroscopy of pumped electrons in quantum Hall regime~\cite{Fletcher2013,Ubbelohde2015,Kataoka2016,Johnson2018}.
The time resolution caused by the geometry is less than \numprint[ps]{20} because hot electrons pumped by tunable-barrier electron pump take around \numprint[ps]{100} to propagate \numprint[\mu m]{5} in quantum Hall edge channel~\cite{Kataoka2016}.

\begin{figure}
	\centering
	\includegraphics[width=8cm]{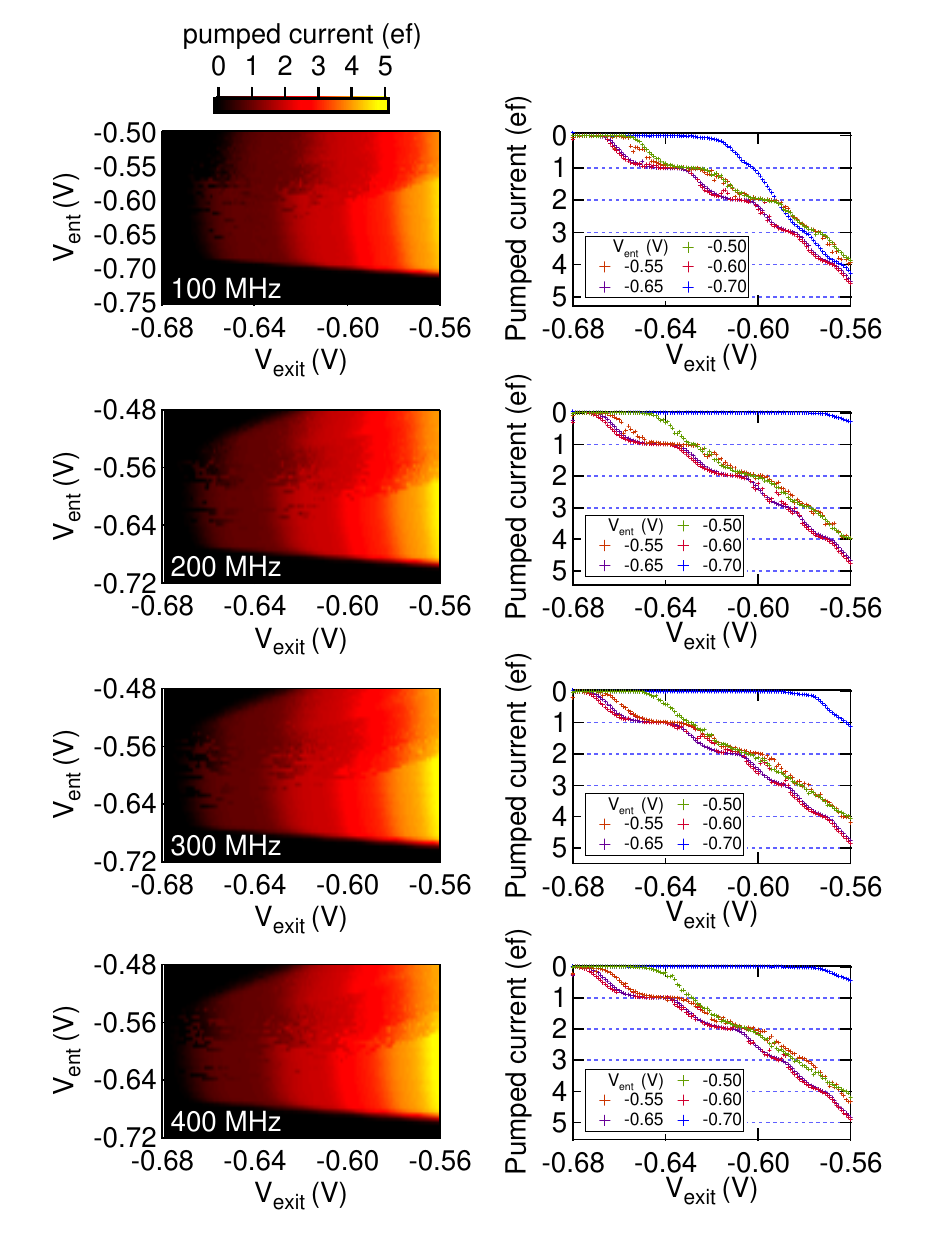}
	\caption{Pumped currents at \numprint[MHz]{100, 200, 300, and 400}. (left) Intensity plot of the current as functions of $V_{\textrm{ent}}$ and $V_{\textrm{exit}}$.
	The electrons are pumped by sine wave whose power is \numprint[dBm]{-4}.
	(right) The line profiles at various entrance gate voltages.}
	\label{fdep}	
\end{figure}
The measurement was done by a homemade liquid helium cryostat ($\sim$ \numprint[K]{4.2}, no magnetic field).
The DC lines of our cryostat contain low pass filters consisting of one \numprint[k\Omega]{1} resistor and two \numprint[nF]{1} capacitors near the sample holder.
The RF line has bias-tees at room temperature to combine the RF signal for periodically tuning the tunneling barrier height and the DC voltage to define the voltage of the ``Entrance gate'' ($V_{\textrm{ent}}$) (see \figref{SEM}).
Only a DC voltage ($V_{\textrm{exit}}$) is applied on ``Exit gate" to define a quantum dot. 
The single electron device formed by the left two gates (colored in orange and green) is used in this paper.
The Entrance gates and Exit gates (see \figref{SEM}) independently pinch off the 2DEG at around \numprint[V]{-0.45}.
A tunable-barrier electron pump operated at relatively high frequency, \numprint[GHz]{1} for example, drops electrons on the source reservoir before the electrons are relaxed to the lower energy state of the quantum dot, which makes it difficult to understand device properties as the single electron source~\cite{Kataoka2011}.
In order to avoid such complications, in the present experiment, the sine wave whose frequency is as low as \numprint[MHz]{100, 200, 300, and 400} is used to operate electrons.
The current generated by the single electron source is measured by digital multi meter (Keithley 2000) followed by the home-made current-voltage converter.

\bfigref{fdep} shows the intensity plots of the current as functions of $V_{\textrm{ent}}$ and $V_{\textrm{exit}}$, and their line profiles at fixed $V_{\textrm{ent}}$ when the power of RF is \numprint[dBm]{-4}~\cite{Blumenthal2007, Kaestner2008APL}.
The pumped current is normalized by elementary charge $e = \numprint[C]{1.602e-19}$ and each own pumping frequency $f$.
The plateaus with the uniform color correspond to where the number of pumped electron per one cycle of RF is fixed.
As clearly shown in the line profiles on each frequency (see the right panels of \figref{fdep}), the plateaus shows the quantization of the multiples of $ef$.
This experimental results successfully prove the desired single electron operation in the device fabricated by our etching technique.


To conclude, we established the techniques to fabricate a narrow 2DEG wire for tunable-barrier electron pump and demonstrated successful single electron operation.
The advantage of the present technique is that EB lithography and photolithography can be done by the same resist in one-step.
Our process is applicable to etching of other materials and dry etching because S1800 series consist of a novolak resin as strong as ZEP520A (ZEON) against dry etching\cite{zep}.
The techniques enable us to omit one etching process and subsequent cleaning process, which improves the quality of the sample.

We thank Jonathan D. Fletcher, Patrick Sea, and Masaya Kataoka in National Physical Laboratory in England for their education on tunable-barrier electron pump during SN's internship in England.
SN was partially supported by Global Internship Program funded by Program for Leading Graduate Schools in Osaka University: INTERACTIVE MATERIALS SCIENCE CADET.
This work is partially supported by JSPS KAKENHI (Grant Numbers JP17J01293, JP18H01815, JP19H00656 and JP19H05826).

The data that support the findings of this study are available from the corresponding author upon reasonable request.


%
%

%



\end{document}